# Derivation of soil-specific streaming potential electrical parameters from hydrodynamic characteristics of partially saturated soils

D. Jougnot[1], N. Linde[1], A. Revil[2,3], and C. Doussan[4]

[1] Institute of Geophysics, University of Lausanne, Lausanne, Switzerland;

[2] Colorado School of Mines, Green Center, Dept of Geophysics, Golden, CO 80401, USA;

[3] ISTerre, CNRS, UMR CNRS 5275, Université de Savoie, 73376 cedex, Le Bourget du Lac, France

[4] EMMAH, UMR 1114, INRA, UAPV, Avignon, France.

Authors contact :

Damien Jougnot: damien.jougnot@unil.ch

Niklas Linde: niklas.linde@unil.ch

André Revil: arevil@mines.edu

Claude Doussan: claude.doussan@avignon.inra.fr

## Abstract

Water movement in unsaturated soils gives rise to measurable electrical potential differences that are related to the flow direction and volumetric fluxes, as well as to the soil properties themselves. Laboratory and field data suggest that these so-called streaming potentials may be several orders of magnitudes larger than theoretical predictions that only consider the influence of the relative permeability and electrical conductivity on the self potential (SP) data. Recent work has partly improved predictions by considering how the volumetric excess charge in the pore space scales with the inverse of water saturation. We present a new theoretical approach that uses the flux-averaged excess charge, not the volumetric excess charge, to predict streaming potentials. We present relationships for how this effective excess charge varies with water saturation for typical soil properties using either the water retention or the relative permeability function. We find large differences between soil types and the predictions based on the relative permeability function display the best agreement with field data. The new relationships better explain laboratory data than previous work and allow us to predict the recorded magnitudes of the streaming potentials following a rainfall event in sandy loam, whereas previous models predict three orders of magnitude too small values. We suggest that the strong signals in unsaturated media can be used to gain information about fluxes (including very small ones related to film flow), but also to constrain the relative permeability function, the water retention curve, and the relative electrical conductivity function.

# 1. Introduction

Under unsaturated conditions, water fluxes are typically inferred from state variables (water content, capillary pressure, or temperature) (e.g. Tarantino et al. 2008, Vereecken et al., 2008). These local and typically disruptive measurements can be complemented with geophysical monitoring and subsequent inversion of geophysical data with a larger support-volume that are sensitive to the above-mentioned state-variables (e.g., Kowalsky et al., 2005). Most of these techniques infer fluxes by data or model differencing in time or space, that is, they are not directly measuring the fluxes occurring at the time of the measurements. The self-potential (SP) method, in which naturally occurring electrical potential differences are measured, provides data that are directly sensitive to water flow (e.g., Thony et al., 1997). The origin of this phenomenon is associated with water flow in a charged porous medium, such as a soil (or more precisely, with the drag of excess charge contained in the diffuse layer in the pore water that surrounds mineral surfaces). The source current density that creates the SP signals has several other possible contributors (e.g., related to redox and diffusion processes), but we focus here on streaming currents, which often tend to dominate in the vadose zone. The generation and behavior of streaming potentials in porous media under two-phase flow conditions have been investigated within an increasing number of publications, but no consensus has been reached concerning how to best model the SP source signals.

Streaming potential responses has been studied at different scales and with different degrees of control (from the field to the laboratory). Thony et al. (1997) were the first to demonstrate experimentally a strong linear relationship between SP signals and water flux in unsaturated soils. Doussan et al. (2002) found based on long-term monitoring in a lysimeter that even if strong linear relationships are present during and after individual rainfall events, no linear relationship can explain data from different soil types and water content conditions. Perrier and Morat (2000) monitored SP signals at an experimental site for one year and proposed a means to explain observed daily variations by considering vadose zone processes. Suski et al. (2006) monitored an infiltration test from a ditch. Using surface-based SP monitoring data from a periodic pumping test, Maineult et al. (2008) observed a clear correlation between pumping and SP signal, but with a time-varying phase lag between the measured SP signals at the ground surface and the in situ pressure heads. This phase lag was explained by Revil et al. (2008) using an hysteretic flow model in the vadose zone. Recently, Linde et al. (2011) showed that SP sources in the vadose zone might strongly influence the measured response in surface-based SP surveys, which has important ramifications as such surveys are often interpreted in terms of groundwater flow patterns only.

Field experiments usually suffer from incomplete knowledge about the variation of relevant variables and boundary conditions with time. It is therefore often necessary to rely on well-controlled laboratory experiments when deriving equations governing streaming potentials under unsaturated conditions. Guichet et al. (2003), Revil and Cerepi (2004), Linde et al. (2007), Revil et al. (2007), Allègre et al. (2010), and Vinogradov and Jackson (2011) have all investigated streaming potentials in the laboratory using either soil or rock samples or 1D column experiments. In addition to low-frequency signals associated with water flow, Haas and Revil (2009) demonstrated the existence of bursts in the electrical field associated with Haines jumps during drainage and imbibition experiments. At an intermediate scale between laboratory and field conditions, Doussan et al. (2002) conducted a six month monitoring experiment of SP signals, pressure, and temperature in a lysimeter under natural conditions (evaporation and rainfall recharge). These authors developed empirical relationships to relate SP measurements and water flux for different rainfall events, but no general relationship was found that could explain all the data.

Different approaches have been invoked to explain and model SP signal generation under unsaturated conditions. Wurmstich and Morgan (1994) proposed an enhancement factor to the saturated streaming potential coupling coefficient equation to model the SP responses to a pumping tests of an oil reservoir. Darnet and Marquis (2004) and Sailhac et al. (2004) introduced Archie's second law in the traditional Helmholtz-Smoluchowki definition of the streaming potential coupling coefficient to account for the partial water saturation, but ignored saturation-induced variations in the relative permeability and excess charge. This theory, like the one proposed by Wurmstich and Morgan (1994), predict an increase of the streaming potential coupling coefficient with decreasing water content, which is in contradiction with laboratory data that generally show decreases with a decreasing water content (among others, Guichet et al., 2003; Revil and Cerepi, 2004; Vinogradov and Jackson, 2011). Revil and Cerepi (2004) explained this behavior in terms of the increased relative importance of surface-related conduction mechanisms with a decreasing water saturation. Saunders et al. (2006) used the model of Revil and Cerepi (2004) to simulate streaming potentials during hydrocarbon recovery. Perrier and Morat (2000) suggested that the streaming potential coupling coefficient should scale with water saturation according to the ratio of relative permeability and relative electrical conductivity. Linde et al. (2007) and Revil et al. (2007) extended this model by suggesting that also the excess charge need to be considered and they scaled it with the inverse of the water saturation. This scaling based on volume averaging is simplified as the volume averaged values are typically very different from the flux-averaged excess charge that influence measured streaming potentials (Linde et al., 2009). Recently, Jackson (2008; 2010) and Linde (2009) proposed models based on a capillary bundle that account for the pore size distribution of partially saturated porous media in the prediction of streaming potentials. The resulting predictions are strongly influenced by both the pore size distribution and the electrical double layer, but no attempts has been made to date to relate these models to available soil-specific hydrodynamic properties. The aim of the present contribution is to propose and test two different models based on soil hydrodynamic properties.

We use the pore size distribution and the excess charge distribution in the Gouy-Chapman layer to derive the effective flux-averaged excess charge density dragged in the medium. The model for each soil type is derived from soil-specific hydrodynamic functions, namely the water retention and the relative permeability functions. For each of these functions, we evaluate for a range of soil textural classes how the effective excess charge in the pore water varies with the effective water saturation. The resulting relationships are then used to determine how the streaming potential coupling coefficient is expected to vary with the effective water saturation. The two approaches are evaluated against the laboratory data of Revil and Cerepi (2004) and the lysimeter monitoring data of Doussan et al. (2002).

## 2 Soil hydrodynamic function-based models

### 2.1 Governing equations and previous work

The two equations that describe the SP response of a given source current density $\mathbf{j}_s$ (A m$^{-2}$) is given by Sill (1983)

$$\mathbf{j} = \sigma \mathbf{E} + \mathbf{j}_S, \qquad [1]$$

$$\nabla \cdot \mathbf{j} = 0, \qquad [2]$$

where $\mathbf{j}$ (A m$^{-2}$) is the total current density, $\sigma$ (S m$^{-1}$) is the bulk electrical conductivity, $\mathbf{E} = -\nabla \varphi$ (V m$^{-1}$) is the electrical field, and $\varphi$ (V) is the electrical potential. The source current densities can be understood as forcing terms that perturb the geological system from electrical neutrality. This induces an electrical current that re-establishes electrical neutrality and the SP response are the associated voltage differences created by this current. In the absence of external source currents it is possible to combine these equations to yield the following governing equation

$$\nabla \cdot (\sigma \nabla \varphi) = \nabla \cdot \mathbf{j}_S.$$ [3]

This partial differential equation can be solved using finite-element or finite-difference techniques given appropriate boundary conditions and exhaustive knowledge about the spatial distribution of $\sigma$ and the source current density $\mathbf{j}_S$ (e.g., Sill, 1983). In the field, the electrical conductivity distribution can be estimated using electrical resistivity tomography (e.g., Günther et al., 2006) or electromagnetic methods (e.g., Everett and Meju, 2005), while the influence of the uncertainty in these models can be evaluated through sensitivity tests (e.g., Minsley, 2007). The focus of this paper is on how to predict $\mathbf{j}_S$ from soil-specific hydrodynamic functions.

Three sources of $\mathbf{j}_S$ may dominate in natural media: electrokinetic processes that are directly related to the water flux in the medium (related to the streaming current density $\mathbf{j}_S^{EK}$), redox processes, and electro-diffusion (see, among others Revil and Linde, 2006). Redox processes can create large SP signals but only under certain restrictive conditions (see discussion in Revil et al., 2009). In the present study, we restrict ourselves to electrokinetic processes that typically dominate in hydrological applications. The water flux follows Darcy's law and can be described by the Darcy velocity $\mathbf{u}$ (m s$^{-1}$) defined by

$$\mathbf{u} = -\frac{k}{\eta_w} \nabla (p_w - \rho_w g z) = -K_w \nabla H,$$ [4]

where $k$ (m$^2$) is the permeability, $\eta_w$ (1.002 × □10$^{-3}$ Pa s at $T$ = 20 °C) is the dynamic viscosity, $p_w$ (Pa) is the water pressure, $\rho_w$ is the water density (1000 kg m$^{-3}$), $g$ is the gravitational acceleration (9.81 m s$^{-2}$), $K_w$ (m s$^{-1}$) is the hydraulic conductivity, and $H$ (m) is the hydraulic head (m). In saturated media, the Darcy velocity is related to the pore water velocity $\mathbf{v}$ (m s$^{-1}$) and the porosity $\phi$ (-) by $\mathbf{u} = \phi \mathbf{v}$.

The streaming current density ($\mathbf{j}_S^{EK}$) is typically described using the streaming potential coupling coefficient $C_{EK}$ (V m$^{-1}$)

$$\mathbf{j}_S^{EK} = \sigma C_{EK} \nabla H,$$ [5]

with $C_{EK}$ defined as

$$C_{EK} = \left. \frac{\partial \varphi}{\partial H} \right|_{\mathbf{j}=0}.$$ [6]

For water-saturated conditions (denoted by superscript *sat*), Revil and Leroy (2004) relate $C_{EK}^{sat}$ to the excess charge in the electrical double layer as

$$C_{EK}^{sat} = \frac{\overline{Q}_v^{sat} k}{\sigma^{sat} \eta_w},$$ [7]

where $\overline{Q}_v^{sat} = \left(1 - f_Q\right) Q_v$ is the excess charge in the Gouy-Chapman layer per pore water volume with $f_Q$ the fraction of excess charge in the Stern layer and $Q_v$ (C m$^{-3}$) the total excess charge that counter balance the mineral surface charges. Equation [7] can be extended for partial saturation in a water-wet media for which we explicitly indicate a dependence of the material properties on the water saturation $S_w$

$$C_{EK}(S_w) = \frac{\overline{Q}_v(S_w)}{\sigma(S_w)} \frac{K_w(S_w)}{\eta_w}.$$ [8]

Note that several functions describing $\sigma(S_w)$ exist in the literature (among other Waxman and Smits, 1968; Rhoades et al., 1989). Laloy et al. (2011) recently published a study investigating the most appropriate pedo-electrical model for a loamy soil.

It is also possible to express $\mathbf{j}_S^{EK}$ at partial saturations as (Revil et al., 2007)

$$\mathbf{j}_S^{EK} = \overline{Q}_v(S_w)\mathbf{u}.$$ [9]

As a first approximation, Linde et al. (2007) and Revil et al. (2007) proposed that $\overline{Q}_v(S_w)$ scales with the inverse of $S_w$, that is,

$$\overline{Q}_v(S_w) = \frac{\overline{Q}_v^{sat}}{S_w}.$$ [10]

Linde (2009) shows that the effective excess charge $\overline{Q}_v^{eff}(S_w)$ dragged in the pore space must be considered as a flux-averaged property that depends on the pore space geometry and the water phase (see also Jackson, 2010). Equation [10] that is based on volume-averaging is therefore only a valid expression for predicting SP signals when $\overline{Q}_v(S_w)$ is evenly distributed throughout the pore space.

In soil hydrology, soil hydrodynamic properties are described by the water retention and the relative permeability function. The first function describes the relationship between the water content, $\theta_w$ (-), (or saturation, $S_w$ (-)) and the matric potential, $h$ (m), whereas the second relates the hydraulic conductivity to the water content. Theoretical formulations of these hydrodynamic properties have often been derived by conceptualizing the soil as a bundle of cylindrical capillaries with a given size density distribution, tortuosity, and connectivity (e.g. Jury et al., 1991).

In the following section 2.2, we describe the electrokinetic behavior and the electrical conductivity of a given capillary. Then in section 2.3 and 2.4 we present two approaches to determine $\bar{Q}_v^{\,eff}(S_w)$ by defining the pore space as a bundle of capillaries that is derived either from the water retention function (i.e., the WR approach) or the relative permeability function (i.e., the RP approach).

## 2.2 Effective excess charge in a capillary

We consider a capillary with a radius $R$ and a length $L_c$. We let $r$ be the distance from the pore wall ($r = 0$) to the center of the capillary ($r = R$). The capillary is saturated by an electrolyte of $N$ ionic species $i$, with concentration $c_i^0$ (mol m$^{-3}$), valence $z_i$ (-), and charge $q_i = e z_i$ (C), where $e$ ($1.6 \times 10^{-19}$ C) is the elementary charge. The ionic strength $I$ (mol m$^{-3}$) of the electrolyte is

$$I = \frac{1}{2}\sum_{i=1}^{N} z_i^2 c_i^0. \qquad [11]$$

Note that the ionic strength is equal to the salinity for binary symmetric 1:1 electrolyte (e.g., NaCl).

We assume—as for silicate and aluminosilicate minerals—that the pore walls have a negative surface charge (the case of positive surface charge can be treated in an analogous manner). To assure electrical neutrality, there exists a balancing excess of cations in the pore water (counterions, while anions are called co-ions). Most of the excess charge is located close to the pore wall in the fixed Stern layer and the remaining part is distributed in the diffuse Gouy-Chapman layer, while the free electrolyte is defined by the absence of excess charge (e.g., Leroy and Revil, 2004). Figure 1a presents a sketch of the charge distribution in the different layers.

The Stern layer contains only counterions (with or without their hydration shell) and its thickness is negligible for typical soils. For example, molecular dynamics simulations in a 0.1 M NaCl–montmorillonite system shows that the thickness of the Stern layer is about 6.1 Å (Tournassat et al., 2009). The interface between the Stern layer and the Gouy-Chapman layer is assumed to correspond to the shear plane, which separates the stationary fluid (due to surface effects) and the moving fluid (see among others, Hunter, 1981; Revil et al., 2002).

The electrical potential along this plane is commonly assumed to correspond to the zeta potential $\zeta$ (V). This potential depends for a given mineral, among other things, on ionic strength, temperature, and pH (e.g., Revil et al., 1999).

The thickness of the Gouy-Chapman layer corresponds roughly to two Debye lengths $l_D$ (Hunter, 1981) defined by

$$l_D = \sqrt{\frac{\varepsilon k_B T}{2 I e^2}},$$ [12]

where $\varepsilon = \varepsilon_r \varepsilon_0$ (F m$^{-1}$) is the pore water permittivity, $k_B$ (1.381 $\times \Box 10^{-23}$ J K$^{-1}$) is the Boltzmann constant, $T$ (K) is the absolute temperature, $\varepsilon_0 = 8.854 \times 10^{-12}$ F m$^{-1}$ is the permittivity of vacuum and $\varepsilon_r$=80.1 at $T$=20°C is the relative permittivity of water. The Gouy-Chapman layer contains distributions of both anions and cations that are linked to the local electrical potential in the pore water $\psi = f(r)$. Pride (1994) expressed for the thin double layer assumption (i.e., the thickness of the double layer is small compared to the pore size) how the local electrical potential depends on the $\zeta$-potential and the distance $r$ from the shear plane as (see also Fig. 2a)

$$\psi(r) = \zeta \exp(-r / l_D).$$ [13]

This equation neglects the effects of the charges of the opposite capillary wall (for the case of overlapping Gouy-Chapman layers, see Gonçalvès et al., 2007), which is a valid assumption in most soils under typical conditions. The counterion and co-ion distributions $c_i = f(r)$ in the pore-water follow (see Fig. 2b)

$$c_i(r) = c_i^0 \exp\left(-\frac{q_i \psi}{k_B T}\right),$$ [14]

where $c_i^0$ is the ionic concentration of $i$ far from the mineral surface (i.e., in the free electrolyte). The excess charge distribution $\bar{Q}_v(r)$ (C m$^{-3}$) in the capillary is (excluding the Stern layer) given by (see Fig. 1b)

$$\bar{Q}_v(r) = N_A \sum_{i=1}^{N} q_i c_i(r),$$ [15]

with $N_A = 6.022 \times 10^{23}$ mol$^{-1}$ being Avogadro's number.

For a laminar flow rate, the velocity distribution $v(r)$ in a capillary of radius $R$ with a given hydraulic head vertical gradient $dh/dz$ is approximated by the Poiseuille model (Fig. 1c)

$$v(r) = \frac{\rho_w g}{4 \eta_w \tau} \left( R^2 - (R-r)^2 \right) \frac{dh}{dz}, \qquad [16]$$

where $\tau$ is the tortuosity of the capillary ($L_c/L$), where $L$ is the length over which the pressure difference is applied. The average velocity $v^R$ (m s$^{-1}$) in the capillary is

$$v^R = \frac{\rho_w g}{8 \eta_w \tau} R^2 \frac{dh}{dz}. \qquad [17]$$

By integration of the flux over the total area of the capillary, one can recover the flux-averaged excess charge, that is, the effective excess charge carried by the water flux in the capillary $\bar{Q}_v^{eff,R}$ (C m$^{-3}$) by

$$\bar{Q}_v^{eff,R} = \frac{\int\limits_{r=0}^{R} \bar{Q}_v(r) v(r) r \, dr}{\int\limits_{r=0}^{R} v(r) r \, dr}. \qquad [18]$$

Figure 1 presents a conceptual view of the electrical double layer model (Fig. 1a), the calculated excess charge distribution using Eq. [15] (Fig. 1b), and the calculated pore fluid velocity using Eq. [16] (Fig. 1c).

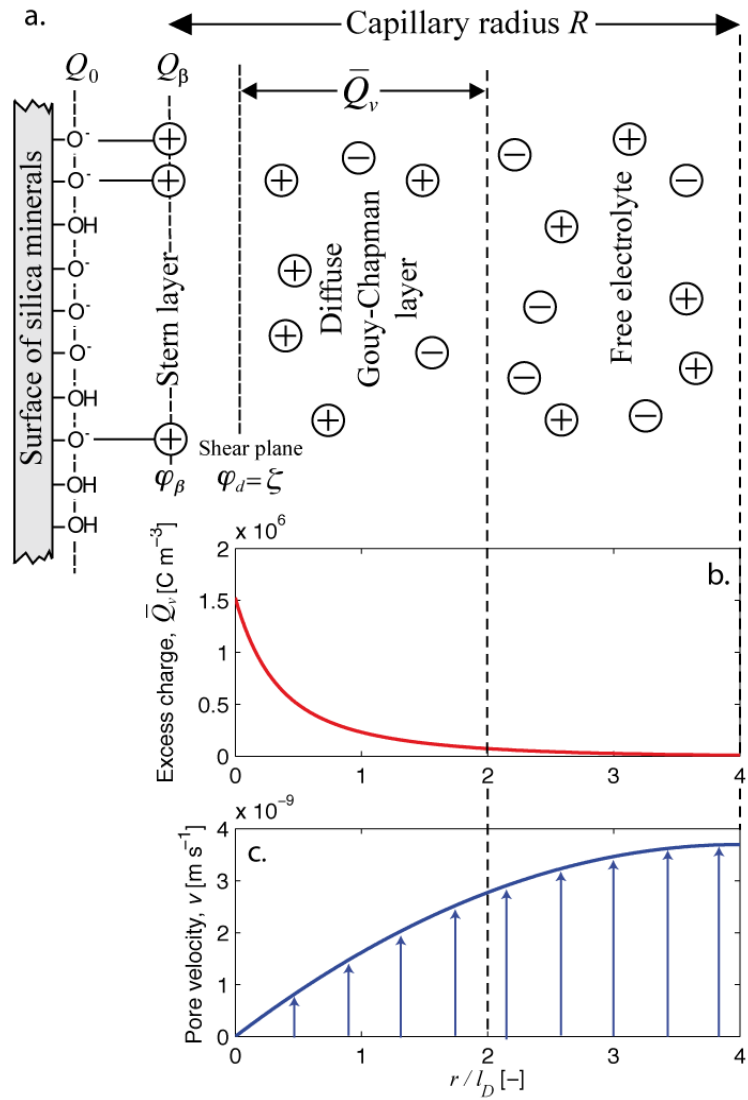

Fig. 1. Distribution of charge and hydraulic flow in a capillary with a radius $R = 4\ l_D$ saturated with a NaCl electrolyte ($10^{-3}$ mol L$^{-1}$, $l_D = 9.70 \times 10^{-9}$ m): (a) sketch of the electrical double layer, (b) excess charge density ($\zeta = -70$ mV) and (c) pore velocity distribution as a function of the distance from the pore wall normalized by the Debye length ($l_D$). The arrows stand for a theoretical flow direction and intensity. Note that the Stern layer is compact and its thickness can often be neglected.

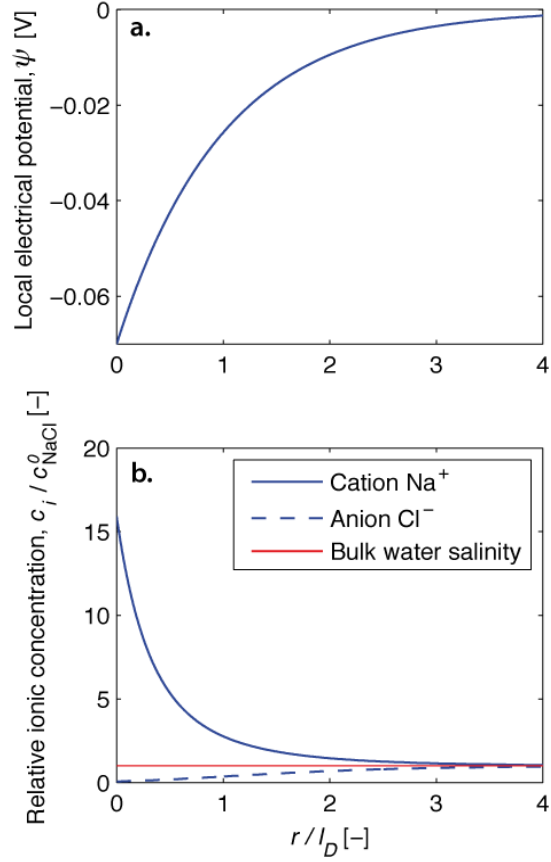

Fig. 2. Effect of surface charge on the pore water properties of a NaCl electrolyte of $10^{-3}$ mol L$^{-1}$ ($l_D = 9.70 \times 10^{-9}$ m) and $\zeta$ = -70 mV (Revil et al., 1999): (a) local potential of the electrical diffuse layer, (b) excess charge density.

## 2.3 From the water retention function to an effective excess charge function

In this section, we express the soil water retention curve in terms of an equivalent bundle of capillaries, which allows us to obtain a relationship between $\bar{Q}_v^{eff}$ and the effective water saturation for a given soil type. The soil water retention function describes the functional relationship between the matric potential (capillary pressure) and water content (or saturation).

The effective water saturation $S_e$ is defined as

$$S_e = \frac{\theta_w - \theta_w^r}{\phi - \theta_w^r},$$  [19]

where $\theta_w = S_w \phi$ (-) is the water content and $\theta_w^r$ (-) is the residual water content after drainage. Van Genuchten (1980) relates $S_e$ to the soil matric potential $h = \dfrac{p_w}{\rho_w g}$ (m) using the following function

$$S_e = \left[ 1 + \left( \alpha_{VG} h \right)^{n_{VG}} \right]^{-m_{VG}}, \qquad [20]$$

where $\alpha_{VG}$ (m$^{-1}$) corresponds to the inverse of the air entry matric potential, while $n_{VG}$ and $m_{VG} = 1 - \dfrac{1}{n_{VG}}$ are curve shape parameters. The air entry matric potential corresponds to the matric potential ($h_e$) at which the soil starts to desaturate.

Another popular water retention function is the one of Brooks and Corey (1964)

$$S_e = \left[ \frac{h_e}{h} \right]^{\lambda_{BC}} \qquad \text{for} \qquad h > h_e, \qquad [21]$$

$$S_e = 1 \qquad \text{for} \qquad h < h_e, \qquad [22]$$

with $h_e$ the air entry pressure (m) and $\lambda_{BC}$ a parameter related to the pore size distribution.

By considering the soil as a bundle of capillaries and applying the Young-Laplace equation, it is possible to relate an equivalent radius $R_j$ (m) to the capillaries $j$ that drain at a specific matric potential by

$$h = \frac{2\gamma \cos \Theta}{\rho_w g R_j}, \qquad [23]$$

where $\gamma$ (0.0727 N m$^{-1}$ at $T$=20°C) is the surface tension of water, $\Theta$ is the contact angle (often considered to be 0°, which yields cos $\Theta$ = 1, see Bear, 1972).

Using Eq. [20 or 21] and Eq. [23] it is thus possible to relate, for a given $S_e$, the size of the capillaries $R_{S_e}$ that drain at an incremental change in $S_e$. This allows us to determine the range and capillary densities of a bundle corresponding to a soil with a given water retention curve. We define the capillary size distribution $f_{WR}(R)$ as

$$\int_{R_{\min}}^{R_{S_e}} f_{WR}(R) \, dR = S_e \left( R_{S_e} \right). \qquad [24]$$

At the scale of the capillary bundle, the electrical formation factor can be expressed under saturated conditions, as

$$\frac{1}{F} = \lim_{\sigma_s \to 0} \left( \frac{\sigma}{\sigma_w} \right) = \phi^m, \qquad [25]$$

where $\sigma_s$ is the surface conductivity and $\sigma_w$ is the electrical conductivity of the pore water, respectively, and $m$ is the cementation index defined by Archie (1942). This exponent is inversely related to the connectivity of the pore space. We assume that the electrical tortuosity under saturated conditions $\tau$=F$\phi = \phi^{1-m}$ also describes the hydrological tortuosity (e.g., Lesmes and Friedman, 2005).

The normalized volumetric flux of water in the pore $v(S_e)$ ($m^3\ s^{-1}\ m^{-2} = m\ s^{-1}$) of the soil can be computed as the sum of the flux of all capillaries up to the size $R_{Se}$

$$v(S_e) = \frac{\int_{R_{min}}^{R_{Se}} v^R(R) f_{WR}(R) dR}{\int_{R_{min}}^{R_{Se}} f_{WR}(R) dR}. \qquad [26]$$

This approach (WR) to calculate $\bar{Q}_v^{eff}(S_e)$ is based on flux-averaging all charges carried by all the capillaries as determined from the water retention curve. We thus define the effective excess charge $\bar{Q}_v^{eff}(S_e)$ as

$$\bar{Q}_v^{eff}(S_e) = \frac{\int_{R_{min}}^{R_{Se}} \bar{Q}_v^{eff,R}(R) v^R(R) f_{WR}(R) dR}{\int_{R_{min}}^{R_{Se}} v^R(R) f_{WR}(R) dR}. \qquad [27]$$

It is then possible to obtain $C_{EK}(S_w)$ by introducing the appropriate $\sigma(S_w)$ function, Eqs. [26] and [27] in Eq. [8]. Note that any hysteretic properties of primarily the water retention function (e.g. Mualem, 1984) but also $\sigma(S_w)$ (Knight, 1991) make the $\bar{Q}_v^{eff}(S_e)$ function hysteretic.

## 2.4 From the relative permeability function to the effective excess charge

In this section, we present an alternative formulation to calculate $\bar{Q}_v^{eff}(S_e)$ that we term the RP approach in which we use the relative permeability function. In this approach, we obtain an equivalent capillary distribution corresponding to a soil with a given relative permeability function that is then used to determine the $\bar{Q}_v^{eff}(S_e)$ relationship.

The relative permeability $k_w^{rel}(S_e)$ is defined as

$$k_w^{rel}(S_e) = \frac{K_w(S_e)}{K_w^{sat}}, \qquad [28]$$

where $K_w(S_e)$ and $K_w^{sat}$, are the partially saturated and the fully saturated hydraulic conductivity ($m\ s^{-1}$), respectively.

Mualem (1976) proposes the following relationship to determine the relative hydraulic conductivity from the soil water retention curve

$$k_w^{rel}(S_e) = S_e^\lambda \left[ \frac{\int_0^{S_e} \frac{dS_e}{h}}{\int_0^1 \frac{dS_e}{h}} \right]^2,$$ [29]

where $\lambda$ is a dimensionless parameter that accounts for hydraulic tortuosity and correlation between pores as a function of $S_e$ (a typical choice is $\lambda = 0.5$). Van Genuchten (1980) introduced his soil water retention function (Eq. [20]) into Mualem's model (Eq. [29]) resulting in the widely used van Genuchten-Mualem (VGM) model

$$k_w^{rel}(S_e) = S_e^\lambda \left[ 1 - \left( 1 - S_e^{1/m_{VG}} \right)^{m_{VG}} \right]^2.$$ [30]

Another popular relative permeability function is the one of Brooks and Corey (1964) that uses a power-law function based on their $\lambda_{BC}$ (Eq. [21]) parameter

$$k_w^{rel}(S_e) = S_e^{\frac{2}{\lambda_{BC}} + 3}.$$ [31]

We now derive a capillary size distribution $f_{RP}(R)$ similarly as for $f_{WR}(R)$ in Eq. [24]. Instead of using the water retention function and Eq. [23], we now use Eq. [17] together with the derivative of the relative permeability function [Eq. 30 or 31] to derive the equivalent $R_{S_e}(S_e)$ that drains at a given $S_e$ as

$$R_{S_e}^2(S_e) = \frac{8\eta_w \tau}{\rho_w g} \frac{K_w^{sat}}{(\phi - \theta_w^r)} \frac{\partial k_w^{rel}(S_e)}{\partial S_e}.$$ [32]

After having determined $f_{RP}(R)$ from $S_e(R_{Se})$, we replace $f_{WR}(R)$ in Eq. [27] to determine the $\bar{Q}_v^{eff}(S_e)$ relationship. One can then recover $C_{EK}(S_w)$ by inserting the resulting $\bar{Q}_v^{eff}(S_e)$ relationship, an appropriate $\sigma(S_w)$ function, and Eq. [32] into Eq. [8].

# 3. Results

## 3.1 Prediction of the relative excess charge and coupling coefficient for a soil data set

We first derive the $\bar{Q}_v^{eff}(S_e)$ relationships of our two approaches using a database of hydrodynamic soil-specific functions (Carsel and Parrish, 1988) compiled from soil water retention measurements of more than 5000 soil samples that are grouped into 12 textural categories. We use the average values of the Van Genuchten parameter ($\phi$, $\theta_w^r$, $\alpha_{VG}$, and $n_{VG}$) and the saturated hydraulic conductivity ($K_w^{sat}$) for each textural category (see Table 1) to calculate the expected $\bar{Q}_v^{eff}(S_e)$ function using the WR (section 2.3) and RP approach (section 2.4).

We hereafter consider the soils presented in Table 1 as being saturated by a NaCl electrolyte at $T = 20°C$ with an ionic strength of $I = 5 \times 10^{-3}$ mol L$^{-1}$. From the empirical relationship proposed by Worthington et al. (1990), this salinity yields a water conductivity equal to $\sigma_w = 0.0603$ S m$^{-1}$. Considering this electrolyte and its concentration, a typical zeta potential at the surface of silica minerals is $\zeta = -61.1$ mV (Revil et al., 1999). We consider hereafter that all the capillary surfaces have this zeta potential.

Table 1. Average values of the Van Genuchten water retention and relative permeability model parameters and saturated hydraulic conductivity of textural soil types (from Carsel and Parrish, 1988).

| Texture [a.] | $\phi$ [-] | $\theta_w^r$ [-] | $\alpha_{VG}$ [m$^{-1}$] | $n_{VG}$ [-] | $K_w^{sat}$ [m s$^{-1}$] | Number of samples [b.] |
|---|---|---|---|---|---|---|
| Clay | 0.38 | 0.068 | 0.8 | 1.09 | $5.56 \times 10^{-7}$ | 333 |
| Clay loam | 0.41 | 0.095 | 1.9 | 1.31 | $7.22 \times 10^{-7}$ | 360 |
| Loam | 0.43 | 0.078 | 3.6 | 1.56 | $2.89 \times 10^{-6}$ | 735 |
| Loamy sand | 0.41 | 0.057 | 12.4 | 2.28 | $4.05 \times 10^{-5}$ | 315 |
| Silt | 0.46 | 0.034 | 1.6 | 1.37 | $6.94 \times 10^{-7}$ | 83 |
| Silt loam | 0.45 | 0.067 | 2.0 | 1.41 | $1.25 \times 10^{-6}$ | 1093 |
| Silty clay | 0.36 | 0.070 | 0.5 | 1.09 | $5.56 \times 10^{-8}$ | 274 |
| Silty clay loam | 0.43 | 0.089 | 1.0 | 1.23 | $1.94 \times 10^{-7}$ | 631 |
| Sand | 0.43 | 0.045 | 14.5 | 2.68 | $8.25 \times 10^{-5}$ | 246 |
| Sandy clay | 0.38 | 0.100 | 2.7 | 1.23 | $3.33 \times 10^{-7}$ | 46 |
| Sandy clay loam | 0.43 | 0.089 | 1.0 | 1.23 | $3.64 \times 10^{-6}$ | 214 |
| Sandy loam | 0.41 | 0.065 | 7.5 | 1.89 | $1.23 \times 10^{-5}$ | 1183 |

a. The textural groups correspond to the USDA classification scheme
b. Average number of samples used to determine the parameters for each soil texture

Figure 3 presents the evolution of the relative excess charge $\bar{Q}_v^{eff,rel}$ (i.e., normalized by the value at full saturation) using the hydrodynamic properties of the various textural classes (Table 1) and the two proposed approaches. Both models predict an important increase of $\bar{Q}_v^{eff,rel}$ with decreasing saturation. This is consistent with the assumption of Linde et al. (2007), but the new models show much stronger increases at low saturations (three to seven orders of magnitudes depending on the soil type and approach used). The WR approach (Fig. 3a) predicts increases of $\bar{Q}_v^{eff,rel}$ that are several orders of magnitudes larger than for the RP approach (Fig. 3b).

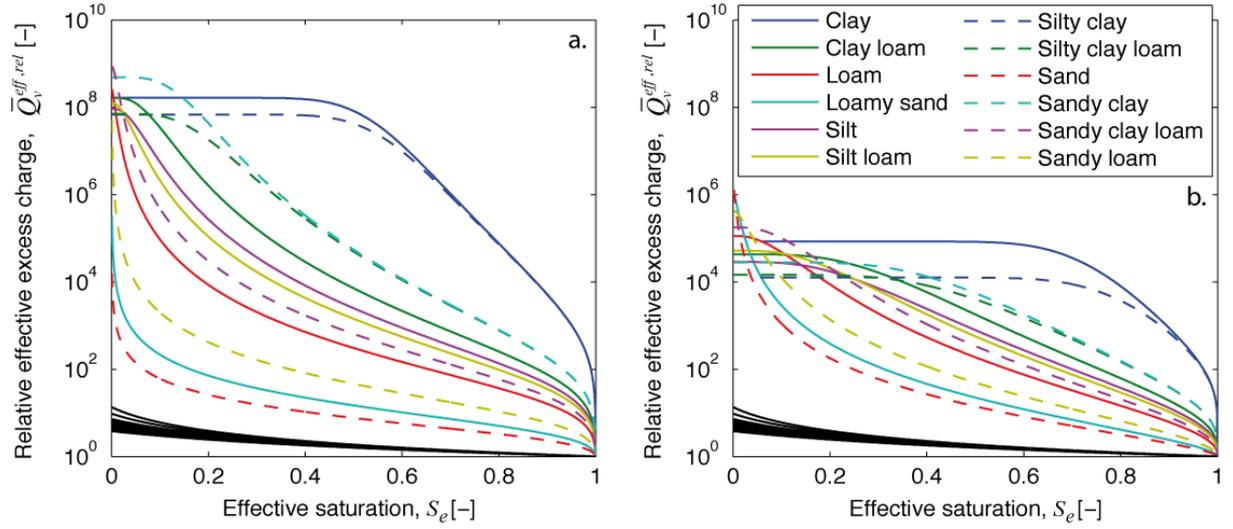

Fig. 3. The $\bar{Q}_v^{eff,rel}(S_e)$ relationships computed by the WR (a.) and RP approaches (b.). The solid black lines correspond to the model of Linde et al. (2007) ($\bar{Q}_v(S_w) = \bar{Q}_v^{sat}/S_w$) for the different soil types (the many neighboring lines arise due to differences in the residual water content among soil types).

Jardani et al. (2007) propose the following empirical relationship between effective excess charge ($\bar{Q}_v^{eff,sat}$) and permeability $k$ (m$^2$) under saturated conditions

$$\log_{10}(\bar{Q}_v^{eff,sat}) = -0.82 \log_{10}(k) - 9.23. \qquad [33]$$

Figure 4 displays the predicted $\bar{Q}_v^{eff,sat}$ as a function of $k$ for the two approaches. For each approach, the predictions closely follow a log-log relationship. The correspondence with the general trend of the experimental data is overall satisfactory, but the absolute values are rather bad for the WR approach (the permeability is over estimated and the $\bar{Q}_v^{eff,sat}$ is underestimated). This is due to the simplification made in the WR approach when computing the permeability directly from the water retention function, while the RP approach use the permeability of Carsel and Parrish (1988) (calculated from $K_w^{sat}$ in Table 1). The resulting linear regression models

$$\log_{10}(\bar{Q}_v^{eff,sat}) = -0.77 \log_{10}(k) - 9.14, \qquad [34]$$

$$\log_{10}(\bar{Q}_v^{eff,sat}) = -0.76\log_{10}(k) - 8.01 , \qquad [35]$$

for the WR and RP approach, respectively, are rather similar to Eq. [33]. Sensitivity tests based on the ionic strength have shown that when $I$ increases, $\bar{Q}_v^{eff,sat}$ decreases (within half an order of magnitude for $I \in \left[10^{-4}; 10^{-1}\right]$ mol L$^{-1}$), but the slope of the log-log relationship remains similar to Eq. [34] and [35]. These results indicate a strong relationship between $\bar{Q}_v^{eff}$ and $k$ through the pore size distribution.

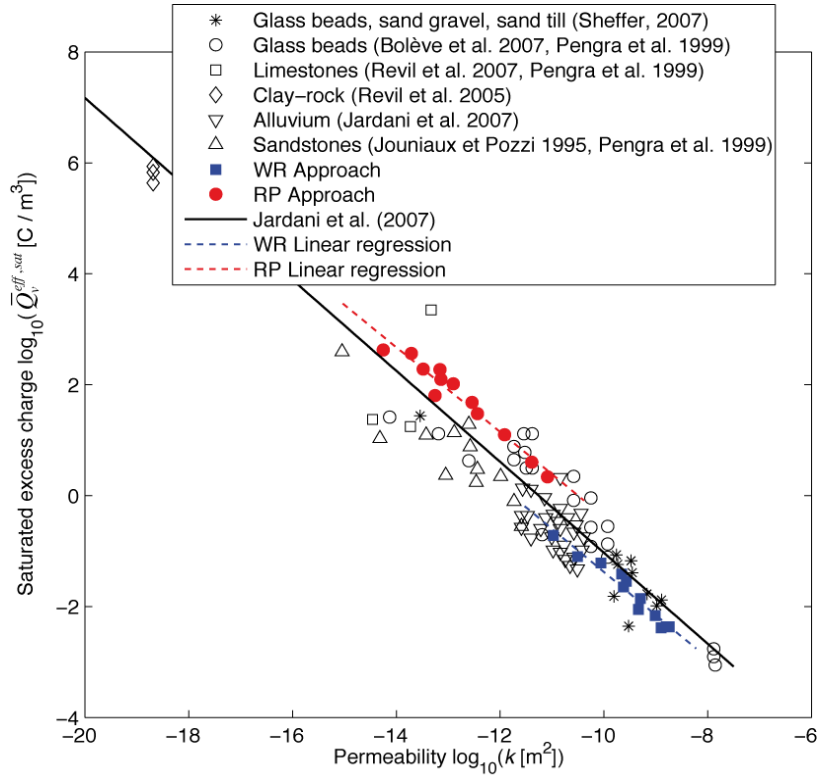

Fig. 4. The predicted effective excess charge using the WR and RP approaches for the different soil textures in saturated conditions. The Jardani et al. (2007) empirical relationship is shown with other data.

Following the proposed approaches, it is possible to predict the evolution of the streaming potential coupling coefficient from three soil specific parameters [Eq. 8]: $\theta_w(h)$, $K_w(h)$, and $\sigma(S_w)$. But, to the best of our knowledge, very few published datasets on soil samples are available that include all three relations. We use the data from Doussan and Ruy (2009) that measured these relationships for: Fontainebleau sand, Collias loam, and Avignon silty clay loam (Fig. 5). As pointed out by the authors, the data cannot be properly described by the traditional water retention and relative permeability functions. We used a cubic interpolation function to describe the parameter evolution with respect to matric potential.

Due to the significant standard deviation of the hydraulic conductivity data, we used the mean as proposed by Doussan and Ruy (2009). We extrapolated the relative permeability up to $h = 10^6$ m based on the last data points and van Genuchten Mualem parameters for corresponding soils.

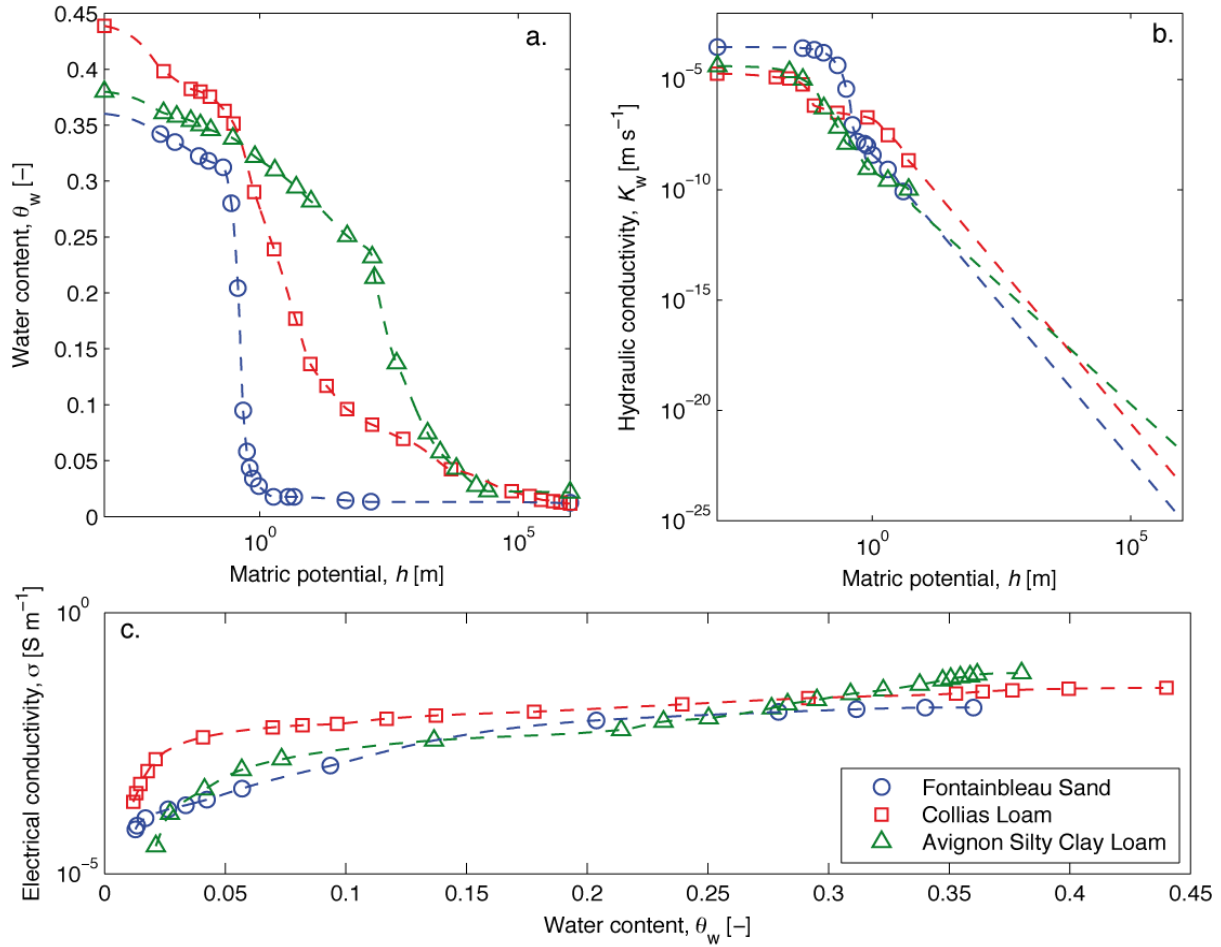

Fig. 5. Properties of three different soil types: (a.) water content and (b.) hydraulic conductivity as a function of matric potential, and (c.) electrical conductivity as a function of water content. Symbols correspond to measurements of Doussan and Ruy (2009), while the dashed lines represent the interpolations of the measurements used in this study.

Figure 6 shows $\bar{Q}_v^{eff,rel}(S_e)$ and $C_{EK}^{rel}(S_e)$ predicted from Eq. [8] using the WR and RP approaches. Figure 6a and 6b show that the predicted excess charges have a similar behavior as for the averaged Carsel and Parrish parameters (Fig. 3), with $\bar{Q}_v^{eff}(S_e)$ varying strongly between soil types. From the predicted $\bar{Q}_v^{eff,rel}(S_e)$, the interpolated $K_w(h)$ (Fig. 5b), and $\sigma(S_w)$ (Fig. 5c), we predicted how the streaming potential coupling coefficient varies with saturation (Fig. 6c and 6d). The behavior of $C_{EK}^{rel}(S_e)$ strongly depends on the different parameters.

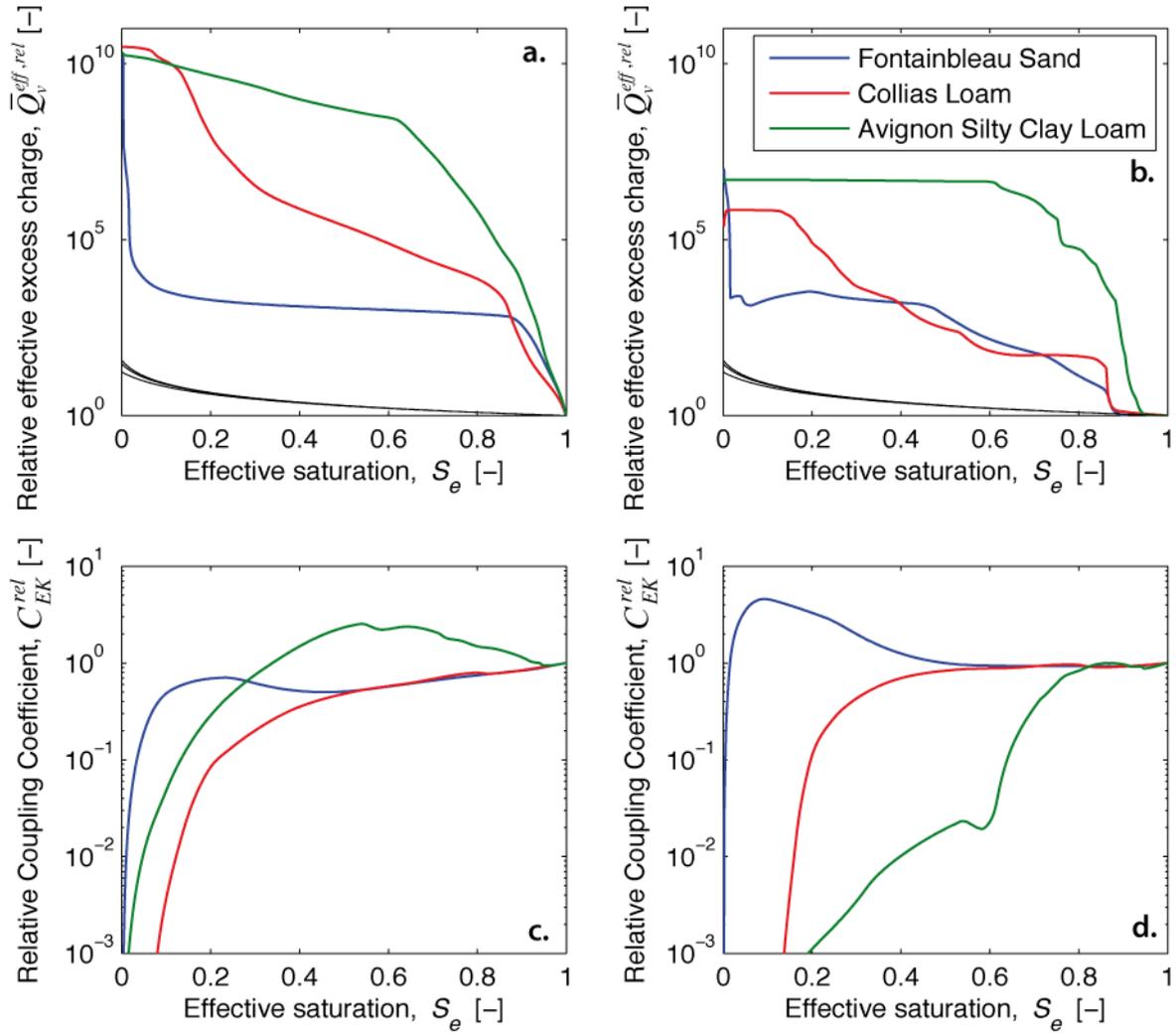

Fig. 6. Relative excess charge (a., b.) and streaming potential coupling coefficient (c., d.) as a function of $S_e$ predicted by the WR and RP approaches, respectively. These predictions have been calculated from the parameter functions showed in Fig. 5. The thin black lines in Fig. 6a. and 6b. correspond to the model of Linde et al. (2007) $\overline{Q}_v(S_w) = \overline{Q}_v^{sat}/S_w$.

## 3.2 Application to laboratory data

We now apply the WR and RP approaches to the laboratory data of Revil and Cerepi (2004). These data include electrical conductivity, capillary pressure and streaming potential coupling coefficient as a function of saturation for two dolomite core samples. The NaCl brine used for the measurements had an ionic strength $I = 8.6 \times 10^{-2}$ mol L$^{-1}$ and a conductivity of $\sigma_w = 0.93$ S m$^{-1}$. For the electrical behavior, Revil and Cerepi (2004) use

Archie's second law to model the relative electrical conductivity $\sigma^{rel} = \sigma(S_w)/\sigma^{sat} = S_w^n$. The hydrological behavior is described using the Brooks and Corey model (Eqs. [21], [22], and [31]). Table 2 presents the parameters used by Revil and Cerepi (2004) to describe the electrical and hydrological properties of the two samples (Figs. 7a and 7b).

Table 2. Electrical and hydrologic parameter values used for the dolomite samples.

| Sample | Porosity | Electrical parameter | | Hydrological parameter | | |
|--------|----------|----------------------|------|------------------------|------|------|
| | $\phi$ [-] [a.] | m [-] [a.] | n [-] [a.] | $S_w^r$ [-] [b.] | $h_e$ [m] [b.] | $\lambda_{BC}$ [-] [b.] |
| E3 | 0.203 | 1.93 [a.] | 2.70 | 0.36 | 2.40 | 0.87 |
| E39 | 0.159 | 2.49 [a.] | 3.48 | 0.40 | 11.52 | 1.65 |

a. From Revil and Cerepi (2004)

b. Parameters fitted from Revil and Cerepi (2004) experimental results

Figure 7c presents the predicted relative streaming potential coupling coefficients using the WR and RP approaches and the predictions of Revil et al. (2007) (see Eq. [10]). The relative streaming potential coupling coefficient predicted from the water retention function (Fig. 7b) fits the E3 sample measurements very well and provide satisfactory values for the E39 sample (Fig. 7c). For all samples, the RP approach tends to overestimate the relative streaming potential coupling coefficient. Note that the relative permeability function is not based on actual measurements, but was derived from the Brooks and Corey (1964) model that is based on the assumption that the $\lambda_{BC}$ describing the water retention function is appropriate to describe the relative permeability function (Eq. [31]). The volume averaging approach of Linde et al. (2007) (Eq. [10]) clearly underestimates $C_{EK}^{rel}(S_w)$ (see also discussion in Allègre et al., 2011). The predicted $\bar{Q}_v^{eff}(S_w)$ from the WR approach is at low saturations several orders of magnitude larger than the predictions of Linde et al. (2007) (e.g., $\bar{Q}_v^{eff}(S_w^r) = 3.4 \times 10^5 \bar{Q}_v(S_w^r)$).

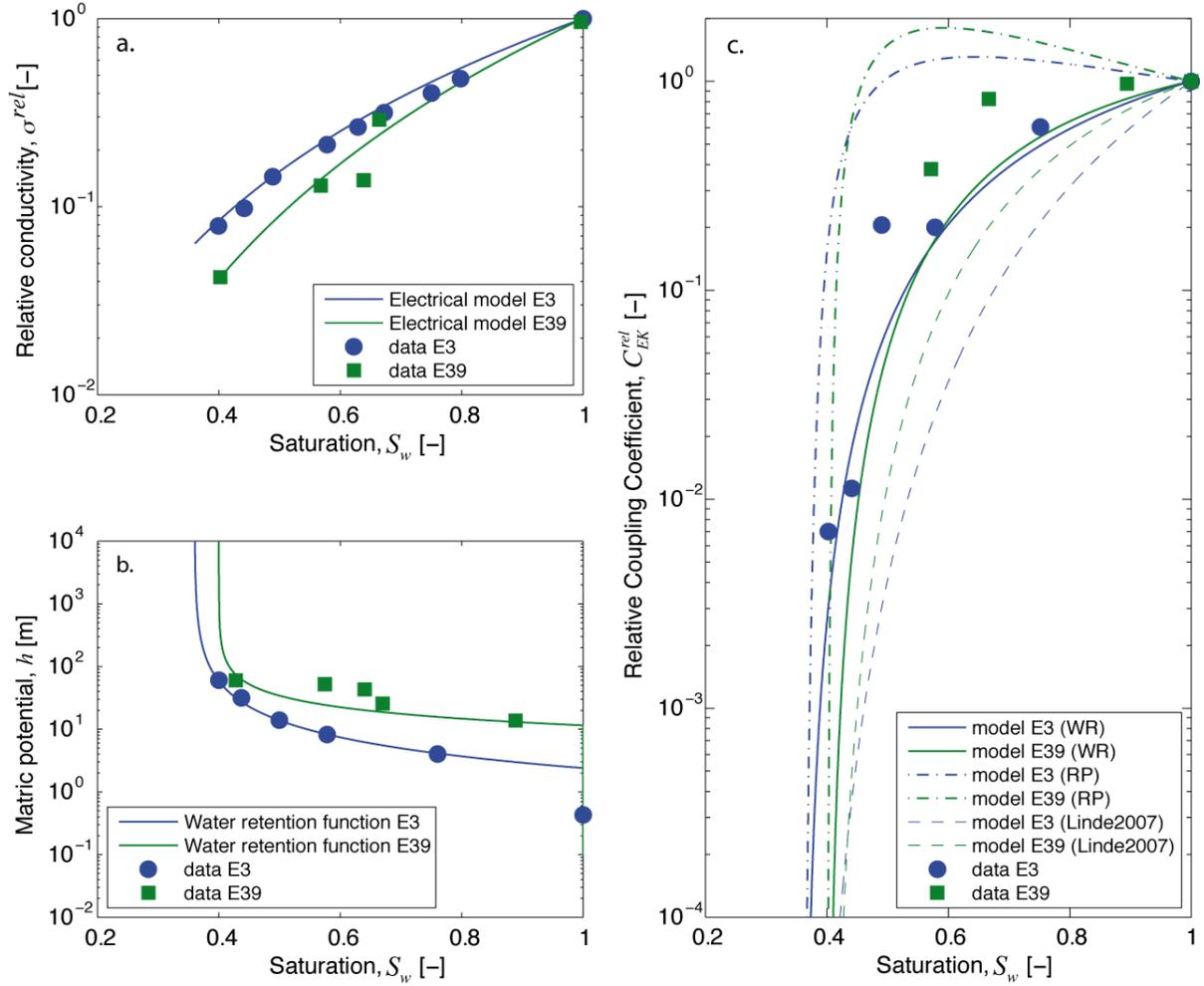

Fig. 7. Application of the proposed approaches to a data set obtained on two dolomite samples from Revil and Cerepi (2004): (a) Relative electrical conductivity, (b) matric potential, and (c) relative streaming potential coupling coefficient versus saturation. The two thin dashed lines in Fig 6c represents the predicted values for the approximation $\overline{Q}_v(S_w) = \overline{Q}_v^{sat} / S_w$ .

### 3.3 Application to a lysimeter experiment

We now apply our model to the experimental data acquired by Doussan et al. (2002) in a lysimeter with a 9 m² surface and a 2 m height located at the INRA experimental field site in Avignon, France. The lysimeter was filled with a local sandy loam and instrumented to monitor unsaturated vertical hydraulic flux. The matric potential was monitored at two depths (30 and 40 cm below ground surface) using two tensiometers for a period of 6 months, while SP data were acquired—at two different locations—between the same two depth intervals using unpolarizable Pb/PbCl₂ electrodes (Petiau, 2000) at a 20 cm distance from the tensiometers. The electrodes located at 30 cm depth were chosen as references. The SP data were corrected for temperature effects following Petiau (2000). The pore water conductivity was measured punctually using suction cups at a depth of 35 cm. The soil cation exchange capacity (*CEC*) of the soil was measured under laboratory conditions using the Metson method (Metson, 1956).

The water retention curve and the relative permeability function of the sandy loam were determined under laboratory conditions using the Wind evaporation method (Tamari et al., 1993). The two hydrodynamic functions could not be adequately fitted using the same van Genuchten parameters (see Table 3 for the individually best fitting van Genuchten parameters).

Table 3. Soil properties of the sandy loam soil of Doussan et al. (2002).

| | $\phi$ [-] | $\theta_w^r$ [-] | $\alpha_{VG}$ [m$^{-1}$] | $n_{VG}$ [-] | $K_w^{sat}$ [m s$^{-1}$] |
|---|---|---|---|---|---|
| Water retention function | 0.44 | 0 | 1.13 | 1.36 | - |
| Relative permeability function | 0.44 | 0 | 0.28 | 1.33 | $1.25 \times 10^{-7}$ |

The electrical behavior of the soil was modeled using the Waxman and Smits (1968) model

$$\sigma = \frac{S_w^n}{F} \left( \sigma_w + \frac{\sigma_s}{S_w} \right),$$ [36]

with parameter values $F = 4.54$, $n = 1.877$ (the saturation index) and $\sigma_s = 0.109$ S m$^{-1}$. Due to rainwater infiltration and evaporation, the water conductivity was changing with time as inferred from the measurements in the suction cups ($\sigma_w \in [0.06; 0.20]$ S m$^{-1}$).

We now test our proposed approaches on rainfall events occurring during the monitoring period. The climatic conditions during the 6 months can be divided into two parts. No major rain event occurred during the first 90 days. Then a series of rainfall events occurred and we chose the five major events at days 91, 100, 107, 119, and 131. Following Doussan et al. (2002), we divide the rainfall events into an infiltration and a drainage phase. The infiltration phase correspond to an increase of the flux as the rainwater reaches the sensors, while the drainage part is characterized by the decrease of both water content and flux. In their interpretation, Doussan et al. (2002) established different relations between the SP signal and the water flux between these two phases.

A first analysis of these data can be done by investigating the evolution of $\bar{Q}_v^{eff}(S_w)$. Considering the lysimeter as a 1D system, Eq. [3] yields $\sigma \dfrac{d\varphi}{dz} = j_s$. Combined with Eq. [9], it is possible to use this relationship to determine the effective excess charge from the measured quantities

$$\bar{Q}_v^{eff}(S_w) = \frac{\sigma(S_w)}{u}\frac{d\varphi}{dz}. \qquad [37]$$

The SP gradient is calculated from the measured SP signals and the spacing between the two electrodes (10 cm). The Darcy velocity $u$ is inferred from the matric potential measurements using the relative permeability function (at the electrode depths) and the electrical conductivity is predicted at the different inferred water saturations using Eq. [36].

Figure 8 compares $\bar{Q}_v^{eff}(S_w)$ calculated by Eq. [37] with the ones predicted by the proposed approaches using the hydrodynamic function parameters of Table 3. We find that the RP approach provides much better results than the WR approach. For the first event, $\bar{Q}_v^{eff}$ is well predicted by the RP model, while the following events present an increasing discrepancy. It is possible that the drying-wetting in the soils could create hysteretic effects that may explain this observation. We also find that the experimentally inferred $\bar{Q}_v^{eff}$ have a similar behavior with respect to $S_w$ for both the infiltration and drainage phases.

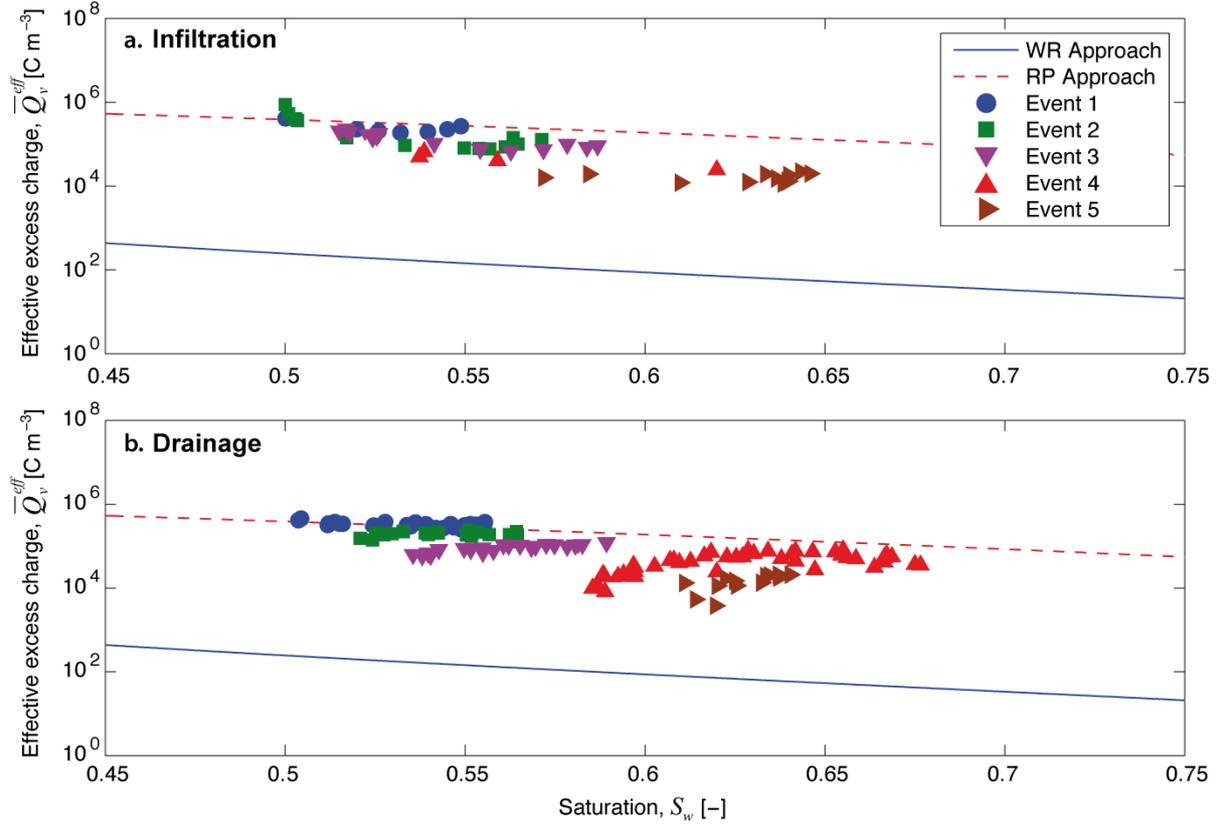

Fig. 8. Effective excess charge as a function of saturation during the five considered rainfall events: (a) the infiltration and (b) the drainage phase. The solid blue and red lines represent the predicted values from the WR and RP approaches, respectively. The different symbols represent the $\bar{Q}_v^{eff}$ calculated from the measured SP data of Doussan et al. (2002) for the different rainfall events (Eq. [37]).

We now compare $\bar{Q}_v^{eff}$ to the measured $CEC$. The total excess charge in the medium $Q_v$ (Stern + Gouy-Chapman layer) can be calculated from the $CEC$ through the following relationship (Waxman and Smits, 1968)

$$Q_v = \rho_S \left( \frac{1-\phi}{\phi} \right) CEC \ . \qquad [38]$$

From the measurements, $CEC = 5.2 \times 10^{-2}$ mol kg$^{-1}$, and considering the typical silica mineral density $\rho_S = 2700$ kg m$^{-3}$, we find $Q_v = 1.66 \times 10^7$ C m$^{-3}$, which is much higher than $\bar{Q}_v^{eff,sat} = 851$ C m$^{-3}$ at saturation estimated from the RP approach. One reason for this discrepancy is that $\bar{Q}_v^{sat} = \left( 1 - f_Q \right) Q_v$, but even in pure clays, which have the highest $f_Q$ values, the typical observed values are $f_Q \in \left[ 0.75 ; 0.99 \right]$ (Leroy and Revil, 2009). This makes us conclude that the flux-averaged $\bar{Q}_v^{eff,sat}$ for this site is smaller than the volume-averaged $\bar{Q}_v^{sat}$ by two to three orders of magnitudes.

### 3.4 Simulation of the SP response to a rainfall event

We now compare the data of Doussan et al. (2002) with simulations of a single rainfall event and the associated modeled SP response. The numerical simulations were conducted using the finite element modeling software COMSOL Multiphysics 3.5 coupled with the scientific computing environment MATLAB. In the simulation, the water flow was computed using Richard's equation with the van Genuchten parameterization (see Table 3 for the parameter values). Note that the residual water content was set to $\theta_w^r = 0.1$ to reach convergence of the hydrological problem at the beginning of the rainfall event. The source current density was calculated from the computed Darcy velocity (Eq. [9]) and the $Q_v^{eff}(S_w)$ was predicted using the RP approach. Considering the low conductivity of the rainwater (2.5 × 10$^{-3}$ S m$^{-1}$), the transport was simulated to also take variations of the pore water conductivity into account. The electrical conductivity model of Waxman and Smits (1968) was used with the parameters of Doussan et al. (2002). The electrical problem (Eq. [3]) was solved at different times to compute the SP signal arising from the hydrological simulation results.

The simulation was performed considering a 2 m high and 0.05 m wide rectangle. The measurement points correspond to the lysimeter experiment (depths of 0.3 and 0.4 m). The geometry was discretized with a mesh with a side length smaller than 30 mm and a mesh refinement down to 5 mm from the surface down to the two measurement points. The hydrological boundary conditions were Neumann boundary conditions on the lateral sides (no water flow), a constant water table at the bottom, and imposed flux at the top (Fig. 9a). The system was assumed to be in hydrostatic equilibrium before the rainfall event with the initial level of the water table determining the water content distribution in the medium. The boundary conditions for the electrical problem were defined as a Neumann condition (electrical insulation) with a reference ($\varphi = 0$ V) at a depth of 0.30 m as in Doussan et al., (2002).

Figure 9 shows the simulation results for different initial water table levels at depths: $WT_{ini} = 4.5, 5.5, 6.5, 7.5,$ and 8.5 m. These values were chosen to represent the range of the experimental hydraulic head of Doussan et al. (2002) prior to rainfall event 1 (day 91). The imposed flux (Figure 9a) at the top corresponds to the rainfall intensity of event 1, which was interpolated from hourly data measured in the vicinity of the lysimeter. Figure 9b shows the variation of the matric potential at a depth of 0.35 m, while the corresponding SP signal between 0.40 and 0.30 m depth is shown in Fig. 9c. The initial level of the water table has clearly a strong influence on the SP response.

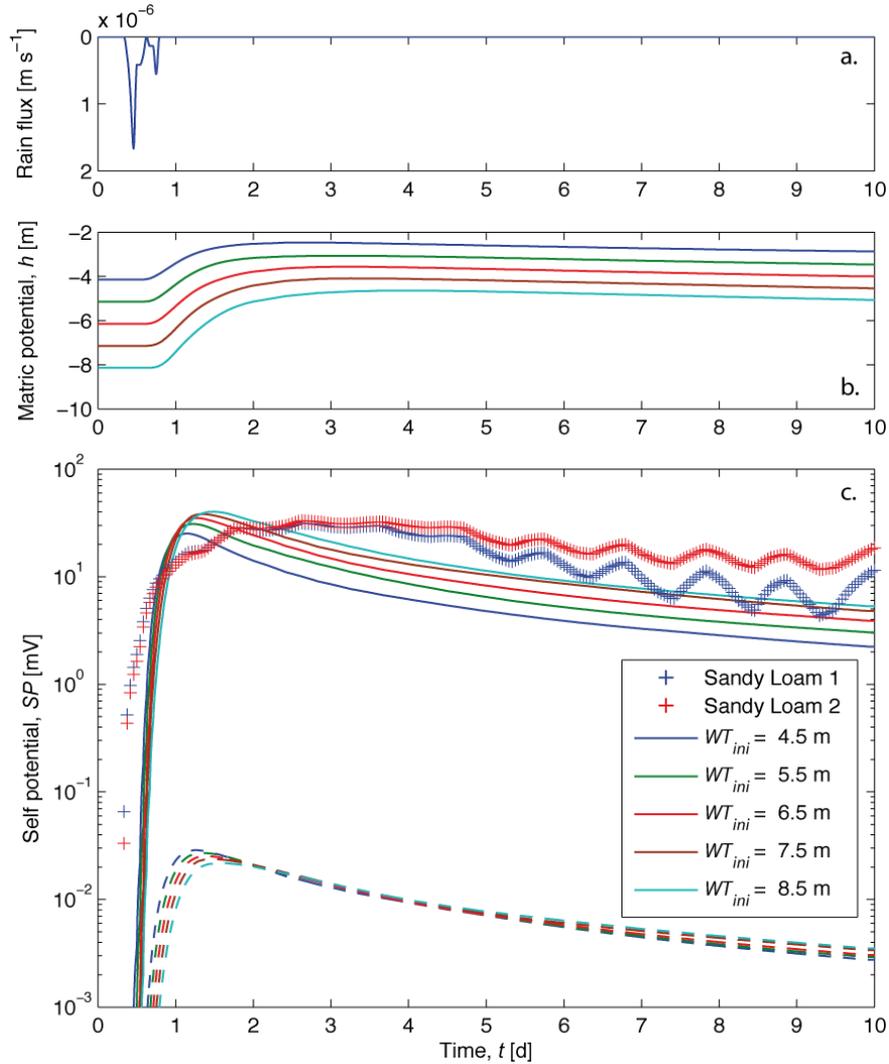

Fig. 9. Predicted SP signals due to rainfall for different initial water table levels: (a) imposed flux from the climatic data of Doussan et al. (2002) for rainfall event 1 (day 91), (b) the modeled matric potential at 35 cm, and (c) the SP signal between 30 and 40 cm depth. Sandy Loam 1 and 2 SP data come from the lysimeter experiment of Doussan et al. (2002). The five dashed lines in Fig 9c represent the predicted SP values for the approximation $\overline{Q}_v(S_w) = \overline{Q}_v^{sat}/S_w$.

The new proposed model explains the experimental data much better than the model of Linde et al. (2007) (dashed lines in Fig. 9c). Considering $WT_{ini} = 6.5$ m, the normalized

*RMS* computed for the model based on the RP approach is 52.3 %, while the signal predicted from the Linde et al. (2007) model has a *RMS* = 97.5 %. We believe that a better description of the initial hydrological conditions would further improve the simulation results of the RP approach. Indeed, it is unlikely to find a hydrostatic equilibrium in a natural soil under in-situ conditions. In addition, evaporation processes were not taken into account in the modeling.

For the 6.5 m deep initial water table, Fig. 10 shows profiles of the simulated SP signal and the matric potential distribution over the first 0.50 m as a function of time (up to 10 days). The SP response due to the rainfall event shows fairly large values at the surface (up 20 mV). The signal peak is followed by a relaxation as the system returns to equilibrium. The relaxation time strongly depends on the water saturation.

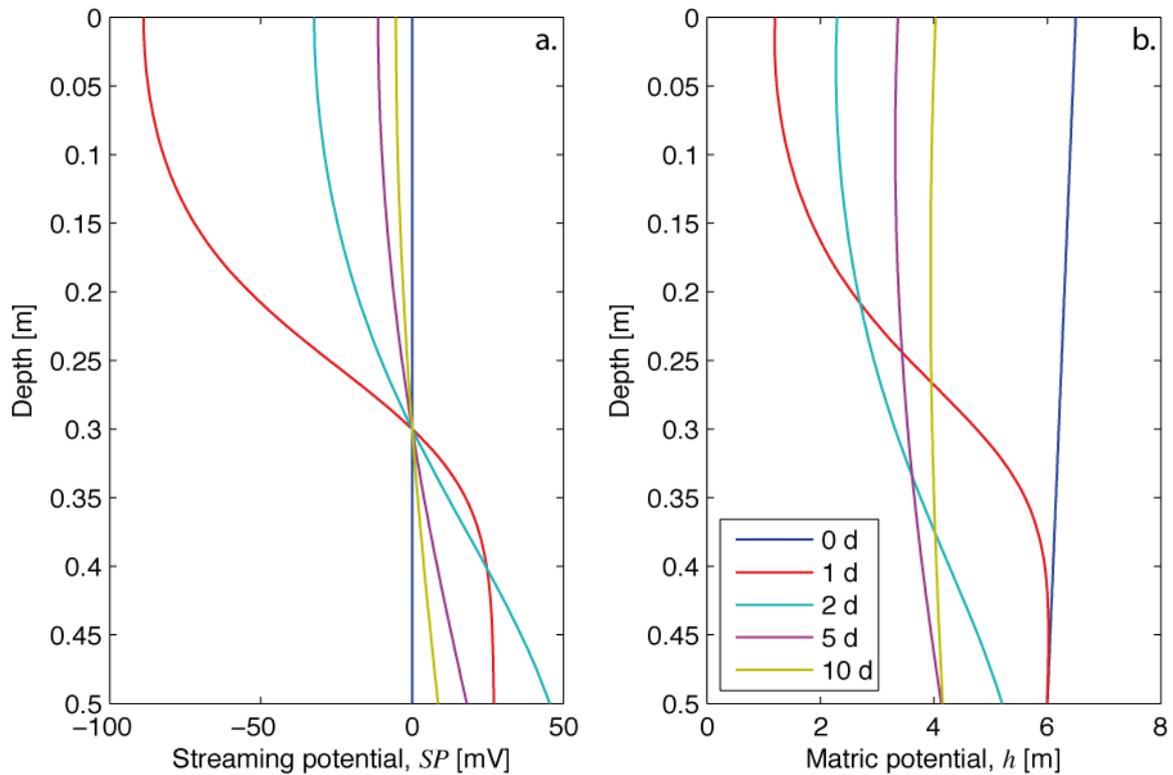

Fig. 10. (a) Streaming Potential signal and (b) matric potential as a function of depth in the medium at 0, 1, 2, 5 and 10 days for a water table at 4.5 m depth. Note that the reference electrode is at 0.3 m depth.

The simulation results give additional confidence in the proposed approach to determine the variation of $\bar{Q}_v^{eff}$ as a function of saturation using hydrodynamic functions. The results also demonstrate that it is possible to obtain fairly large SP signals within a partially saturated medium even when $C_{EK}^{rel}(S_w)$ decreases with $S_e$ as the gradients in hydraulic head can be very large, for example, due to a perturbation of an initially dry soil by a rainfall.

## 4. Discussion

The two new approaches to predict soil-specific flux-averaged $\bar{Q}_v^{eff}(S_e)$ have a higher predictive capacity than the model of Linde et al. (2007) and Revil et al. (2007). The primary reason for this improvement is that the volume-averaging used in the latter approach is based on the assumption of a uniform distribution of excess charge in the pore space. As shown in Fig. 9, the predictions based on $\bar{Q}_v(S_w) = \bar{Q}_v^{sat}/S_w$ is smaller than the measured SP signals by 3 to 4 orders of magnitude, while the RP approach provides values in the same order of magnitude as the field data. This improvement is achieved by considering a more complex model of $\bar{Q}_v^{eff}(S_e)$ derived from hydrodynamic functions that explicitly considers that $\bar{Q}_v^{eff}(S_e)$ is a flux-averaged property. We find that the RP approach is more reliable than the WR approach, which nevertheless provide rather good results in terms of relative variations with respect to water saturation. The better performance of the RP approach is likely caused by an improved inference of the equivalent pore size distribution compared with the WR approach.

The numerical simulations highlight that the SP signals are strongly related to the distribution of excess charge in the pore space and the velocity distribution in the pore space (Fig. 1). The very important difference between the flux-averaged $\bar{Q}_v^{eff}$ calculated from experimental results (Fig. 8) and the volume-averaged excess charge $Q_v$ determined from $CEC$ measurements (see section 3.3) is due to that a disproportionately large fraction of fluid flow takes place outside of the diffuse Gouy-Chapman layer.

The results presented here explain how SP signals can significantly increase at low saturation even if the streaming potential coupling coefficient tends to decrease with saturation. This happens as the hydraulic head gradients in the unsaturated zone can be very large and it is the combined effects of the coupling coefficient and the hydraulic head that creates the SP signal for a 1-D system. Another explanation is offered by rewriting Eq. [37] as

$$\nabla \varphi = \frac{\bar{Q}_v^{eff}(S_w)}{\sigma(S_w)} u.$$

[39]

This equation is as Eq. [37] only valid under 1-D conditions. At low water saturations we found (Fig. 3) that the increase in $\bar{Q}_v^{eff}(S_w)$ might be larger than 1000 compared with saturated conditions and there is nothing unusual about $\sigma(S_w)$ decreasing with a factor of 10 at lower saturations. This means that SP signals at low saturation might be as large as for saturated conditions even when the flux is $10^{-4}$ times smaller.

Equation [39] also highlights why it is unrealistic to expect a linear relationship between SP values and water flux over a large saturation range. Doussan et al. (2002) found that the linear regression models between SP values and water flux could be improved by considering the water content. They attribute this to variations in electrode contact, but is more likely caused by improved linearizations of the underlying non-linear relationship. The difference in the predictions compared with volume averaging is about 100 at low water saturations. This discrepancy explains why Linde et al. (2011) could not simulate the observed SP magnitudes observed on a gravel bar following rainfall. The increased sensitivity to water flow under unsaturated conditions might explain the slow and often incomplete relaxation of SP signals following drainage (e.g. Allègre et al., 2010).

Our findings open up exciting possibilities of using the SP method to monitor very small flows at low saturations, such as those due to evaporation. This would necessitate co-located measurements of bulk electrical conductivity, water saturation, and a good description of hydrodynamic soil properties.

## 5. Conclusions

Soil-specific water retention and relative permeability functions together with a relative conductivity function are needed to predict the streaming potential coupling coefficient under unsaturated conditions and hence SP signals. Most previous studies has ignored or severely underestimated the importance of accurately modeling the scaling of the effective excess charge, which we here predict from the above-mentioned soil-specific hydrodynamic functions. Using a capillary tube model, we find that the effective (flux-averaged) excess charge is for typical soils two-to-three orders of magnitude larger than volume-averaged estimates, which translate to equally larger SP signals. The improvement with respect to existing theory is demonstrated against laboratory data and by comparing the modeled SP response caused by precipitation on a sandy loam with field data. For this data set, the initial water content, the water retention and relative permeability, as well as relative conductivity function was independently measured in the laboratory. The new theory predicts both the right magnitudes and the slow relaxation of the observed SP signal, while this was not possible using volume averaging. It is of course an advantage to have access to laboratory or in situ measurements of hydrodynamic functions, but the presented predictions based on the Carsel and Parrish (1988) database provide a rather good idea about the expected variations of $\bar{Q}_v^{eff,rel}(S_w)$ for different soil types.

Our work provide a credible explanation for the often surprisingly large SP signals that are observed at low water saturation and it opens up the perspective of using SP signals to characterize film flow and evaporation processes. It also suggests that SP signals in the vadose zone can become a useful data source when estimating fluxes (or at least flux directions) in the unsaturated zone and for inverse modeling applications. It also highlights the importance of considering vadose zone processes in general SP surveys as flows that are $10^{-4}$-$10^{-5}$ times smaller than under saturated conditions may in dry soils lead to SP gradients of the same magnitude.

## Acknowledgments

We would like to thank Dani Or for some very useful suggestions in the early stages of this research. Financial support from Fondation Herbette is gratefully acknowledged. We thank Associate Editor Alex Furman and the three anonymous reviewers for detailed comments that helped us to improve the paper.

**Notations**

| Symbol | Description | Units |
|---|---|---|
| | *Electrical and electrochemical variables* | |
| $c_i^0$ | Ionic concentration of the species $i$ in the free electrolyte | mol m$^{-3}$ |
| $c_i$ | Ionic concentration of species $i$ | mol m$^{-3}$ |
| $C_{EK}$ | Streaming potential coupling coefficient | V Pa$^{-1}$ |
| $CEC$ | Cation Exchange Capacity | mol kg$^{-1}$ |
| $\mathbf{E}$ | Electrical field | V m$^{-1}$ |
| $f_Q$ | Fraction of the excess charge in the Stern layer | - |
| $F$ | Electrical formation factor | - |
| $I$ | Ionic strength | mol m$^{-3}$ |
| $\mathbf{j}$ | Total current density | A m$^{-2}$ |
| $\mathbf{j}_s$ | Source current density | A m$^{-2}$ |
| $l_D$ | Debye length | m |
| $m$ | Electrical cementation index | - |
| $n$ | Electrical saturation index | - |
| $N$ | Number of ionic species $i$ | - |
| $q_i$ | Charge of the ionic species $i$ | C |
| $Q_v$ | Volumetric excess charge | C m$^{-3}$ |
| $\overline{Q}_v$ | Volume averaged excess charge | C m$^{-3}$ |
| $\overline{Q}_v^{eff}$ | Effective volumetric excess charge | C m$^{-3}$ |
| $T$ | Temperature | K |
| $z_i$ | Ionic valence of the species $i$ | - |
| $\varepsilon$ | Dielectric permittivity | F m$^{-1}$ |
| $\varepsilon_r$ | Relative dielectric permittivity of water | - |
| $\omega$ | Angular frequency | Hz |
| $\varphi$ | Electrical potential of the porous medium | V |
| $\psi$ | Local electrical potential in the pore water | V |
| $\sigma$ | Electrical conductivity of the medium | S m$^{-1}$ |
| $\sigma_w$ | Water electrical conductivity | S m$^{-1}$ |
| $\sigma_S$ | Surface electrical conductivity | S m$^{-1}$ |
| $\zeta$ | Zeta potential | V |

**Hydrological variables**

| | | |
|---|---|---|
| $h$ | Soil matric potential | m |
| $h_e$ | Air entry matric potential | m |
| $H$ | Hydraulic head | m |
| $k$ | Permeability | m$^2$ |
| $K_w$ | Hydraulic conductivity | m s$^{-1}$ |
| $L$ | Medium length | m |
| $L_c$ | Capillary length | m |
| $m_{VG}$ | Van Genuchten curve shape parameter | - |
| $n_{VG}$ | Van Genuchten curve shape parameter | - |
| $p_w$ | Water pressure | Pa |
| $r$ | Distance from the porewall in a given capillary | m |
| $R$ | Radius of a given capillary | m |
| $R_{Se}$ | Equivalent capillary radius that drains a certain saturation | m |
| $S_e$ | Effective water saturation | - |
| $S_w$ | Water saturation | - |
| $\mathbf{u}$ | Darcy velocity | m s$^{-1}$ |
| $\mathbf{v}$ | Pore water velocity | m s$^{-1}$ |
| $\alpha_{VG}$ | Van Genuchten inverse of the air entry pressure | m$^{-1}$ |
| $\gamma$ | Surface tension of the water | N m$^{-1}$ |
| $\lambda$ | Hydraulic tortuosity parameter as a function of saturation | - |
| $\lambda_{BC}$ | Brooks and Corey pore size parameter | - |
| $\eta_w$ | Dynamic viscosity of the water | Pa s$^{-1}$ |
| $\phi$ | Porosity | m$^3$ m$^{-3}$ |
| $\rho_w$ | Water density | kg m$^{-3}$ |
| $\theta_w$ | Volumetric water content | m$^3$ m$^{-3}$ |
| $\theta_w^r$ | Residual water content | m$^3$ m$^{-3}$ |
| $\Theta$ | Contact angle | ° |
| $\tau$ | Tortuosity of the media | - |

| Sub/super-script | Description |
|---|---|
| BC | Relative to the Brooks and Corey model |
| EK | Relative to ElectroKinetic phenomena |
| i | Relative to a given ionic species |
| j | Relative to a given capillary radius |
| rel | Value of a parameter relatively to its value at saturation |
| R | Value for a given capillary with a radius equal to R |
| RP | Relative to the Relative Permeability approach |
| sat | Parameter at saturation ($S_w = 1$) |
| S | Relative to the solid/surface |
| VG | Relative to the van Genuchten model |
| w | Relative to the water phase |
| WR | Relative to the Water Retention approach |

| Physical constants | Description | Value |
|---|---|---|
| $e$ | Elementary charge | $1.6 \times 10^{-19}$ C |
| $g$ | Gravitational acceleration | $9.82$ m s$^{-2}$ |
| $k_B$ | Boltzmann constant | $1.381 \times 10^{-23}$ J K$^{-1}$ |
| $N_A$ | Avogadro's number | $6.022 \times 10^{23}$ mol$^{-1}$ |
| $\varepsilon_0$ | Dielectric permittivity of vacuum | $8.854 \times 10^{-12}$ F m$^{-1}$ |